\documentclass[prl,twocolumn,showpacs,floatfix,amsfonts]{revtex4}
\usepackage{graphicx,graphics,color,epsfig}
\usepackage{bm}
\usepackage{amsmath}
\usepackage{amssymb}

\begin{document}
\preprint{}

\title{The temperature dependence of the local tunnelling conductance in
cuprate superconductors with competing AF order}
\author{Hong-Yi Chen and C.S. Ting}
\affiliation{Texas Center for Superconductivity and Advanced Material,
and Department of Physics, University of Houston, Houston, TX 77204}

\begin{abstract}
Based on the $t-t'-U-V$ model with proper chosen parameters for describing
the cuprate superconductors, it is found that near the optimal doping at low
temperature ($T$), only the pure $d$-wave superconductivity ($d$SC) prevails
and the antiferromagnetic (AF) order is completely suppressed. At higher $T$,
the AF order with stripe modulation and the accompanying charge order may
emerge, and they could exist above the $d$SC transition temperature. We
calculate the local differential tunnelling conductance (LDTC) from the local
density of states (LDOS) and show that their energy variations are rather
different from each other as $T$ increases. Although the calculated
modulation periodicity in the LDTC/LDOS and bias energy dependence of the
Fourier amplitude of LDTC in the "pseudogap" region are in good agreement
with the recent STM experiment [Vershinin $et\; al.$, Science {\bf 303}, 1995
(2004)], we point out that some of the energy dependent features in the LDTC
do not represent the intrinsic characteristics of the sample.

\end{abstract}


\maketitle

Recently, STM or the local differential tunnelling conductance (LDTC)
measurement by Vershinin $et\; al.$ \cite{vershinin303} on slightly
underdoped BSCCO indicated that the electronic states at low energies and at
a temperature $T$ higher than the superconducting transition temperature
($T_c$) in the pseudogap region exhibit an energy-independent spatial
modulation which resulting to a checkerboard pattern with incommensurate
periodicity $4.7a \pm 0.2$ ($a$ is the lattice constant). At very low
temperature, however, no such pattern has been detected \cite{vershinin303},
in agreement with previous measurements \cite{hoffman297, mcelroy422}. In
addition, the Fourier amplitude of the LDTC at the modulation wave-vector
which corresponds the periodicity $4.7a$ increases its magnitude first and
then flattens out as the bias energy decreases. How to understand these
behaviors are outstanding questions which have not been addressed in the
existing literatures. In the present paper we are trying to explain these
issues by adopting the idea of the possible coexistence of the $d$-wave
superconductivity ($d$SC) with antiferromagnetic (AF) order for cuprate
superconductors \cite{inui37, giamarchi43, inaba257, martin56}, and to
examine the formation of the AF order and the accompanying charge order at
finite temperature. The phenomenological $t-t'-U-V$ model will be applied to
describe the cuprate superconductors. With proper chosen parameters, we show
that at low temperature only $d$SC prevails in our system and the AF order is
completely suppressed. The local density of states (LDOS) and LDTC images are
featureless. At higher temperature, it is found that the AF order with stripe
modulation, which is also referred to the spin density wave (SDW), and the
accompanying charge order or the charge density wave (CDW) may show up and
they could even persist at temperatures above the BCS superconducting
transition temperature $T_c^{BCS}$. In the presence of SDW, we show both of
the LDOS/LDTC images to have energy-independent stripe modulation with
spacing of $5a/4a$ spreading over a $48a \times 24a$ lattice. According to
the Fourier analysis of the LDOS images, an average periodicity $4.8a$ could
be assigned for the stripe modulation. If both of the doubly degenerate $x$-
and $y$- oriented stripes have the probability to appear in the time interval
of the measurement or the proximity effect \cite{howald67} exists between
neighboring domains with differently oriented stripe modulations, the
combined LDOS images would have a checkerboard pattern of $4.8a \times 4.8a$
structure. All these features are consistent with the experiment of Vershinin
$et\;al.$ \cite{vershinin303}. In order to compare with the energy variation
of the STM measurements at finite temperature ($T$), the LDTC is needed and
it can be obtained from the LDOS by using the method of convolution. although
both of the energy variations of LDTC and LDOS exhibit the "pseudogap"-like
characteristics \cite{kugler86}, their behaviors at higher temperature are
quite different. While the size of the gap for quasiparticle excitations as a
function of $T$ is measured by the separation between the coherent peaks in
the LDOS, the use of the LDTC or the experimental data to directly determined
this quantity could be misleading.

To model these observed phenomena, we employ an effective
mean-field $t-t'-U-V$ Hamiltonian by assuming that the on-site
repulsion $U$ is responsible for the competing antiferromagnetism
and the nearest-neighbor attraction $V$ causes the $d$-wave
superconducting pairing

\begin{eqnarray}
{\bf H}&=&-\sum_{{\bf ij}\sigma} t_{\bf ij} c_{{\bf
i}\sigma}^{\dagger}c_{{\bf j}\sigma}
+\sum_{{\bf i}\sigma} ( U\langle n_{{\bf i}\bar{\sigma}}\rangle - \mu )
c_{{\bf i}\sigma}^{\dagger}c_{{\bf i}\sigma} \nonumber \\
&&+\sum_{\bf ij} (\Delta_{\bf ij} c_{{\bf i}\uparrow}^{\dagger}
c_{{\bf j}\downarrow}^{\dagger} +\Delta_{\bf ij}^{*} c_{{\bf
j}\downarrow} c_{{\bf i}\uparrow} )\;,
\end{eqnarray}
where $t_{\bf ij}$ is the hopping integral, $\mu$ is the chemical
potential, and $\Delta_{\bf ij}=\frac{V}{2}\langle c_{{\bf i}\uparrow}
c_{{\bf j}\downarrow}-c_{{\bf i}\downarrow}c_{{\bf j}\uparrow}\rangle$ is
the spin-singlet $d$-wave bond order parameter. The Hamiltonian above
shall be diagonalized by using Bogoliubov-de Gennes' (BdG) equations,
\begin{eqnarray}
\sum_{\bf j}^N \left(\begin{array}{cc}
 {\cal H}_{{\bf i}j\sigma} & \Delta_{\bf ij} \\
 \Delta_{\bf ij}^* & -{\cal H}_{{\bf ij}\bar{\sigma}}^*
 \end{array}\right)
 \left(\begin{array}{c}
     u_{{\bf j}\sigma}^n \\
     v_{{\bf j}\bar{\sigma}}^n
 \end{array}\right)
 = E_n
 \left(\begin{array}{c}
     u_{{\bf i}\sigma}^n \\
     v_{{\bf i}\bar{\sigma}}^n
 \end{array}\right)\;,
\end{eqnarray}
where ${\cal H}_{{\bf ij}\sigma}=-t_{\bf ij} + ( U\langle n_{{\bf
i}\sigma}\rangle - \mu ) \delta_{\bf ij}$. Here, we choose the
nearest-neighbor hopping $\langle t_{\bf ij}\rangle = t=1$ and the
next-nearest-neighbor hopping $\langle t_{\bf ij}\rangle = t'=-0.25$ to
match the curvature of the Fermi surface for most cuprate superconductors
\cite{norman63}. The exact diagonalization method to self-consistently
solve BdG equations with the periodic boundary conditions is employed to
get the $N$ positive eigenvalues $(E_n)$ with eigenvectors $(u_{{\bf
i}\uparrow}^n , v_{{\bf i}\downarrow}^n)$ and $N$ negative eigenvalues
$(\bar{E}_n)$ with eigenvectors $(-v_{{\bf i}\uparrow}^{n*} , u_{{\bf
i}\downarrow}^{n*} )$. The self-consistent conditions are
\begin{eqnarray}
\langle n_{{\bf i}\uparrow} \rangle &=&
  \sum_{n=1}^{2N}\left|{\bf u}_{\bf i}^n\right|^2 f(E_n)\;,\;
\langle n_{{\bf i}\downarrow} \rangle =
  \sum_{n=1}^{2N}\left|{\bf v}_{\bf i}^n\right|^2 [1-f(E_n)]\;,
     \nonumber \\
\Delta_{\bf ij} &=& \sum_{n=1}^{2N} \frac{V}{4} ({\bf u}_{\bf i}^n {\bf
v}_{\bf j}^{n*} + {\bf v}_{\bf i}^{n*} {\bf u}_{\bf j}^n) \tanh
(\frac{\beta E_n}{2})\;,
\end{eqnarray}
where ${\bf u}_{\bf i}^n = (-v_{{\bf i}\uparrow}^{n*}, u_{{\bf
i}\uparrow}^n )$ and ${\bf v}_{\bf i}^n = (u_{{\bf i}\downarrow}^{n*},
v_{{\bf i}\downarrow}^n )$ are the row vectors, and $f(E)=1\slash(e^{\beta
E}+1)$ is Fermi-Dirac distribution function. Since the calculation is
performed near the optimally doped regime, the filling factor,
$n_f=\sum_{{\bf i}\sigma} \langle c_{{\bf i}\sigma}^\dagger c_{{\bf
i}\sigma} \rangle /N_xN_y$, is fixed at $0.85$, i.e., the hole doping
$\delta=0.15$. Each time when the on-site repulsion $U$ or the
temperature is varied, the chemical potential $\mu$ needs to be adjusted .

\begin{figure}[t]
\centerline{\epsfxsize=8.0cm\epsfbox{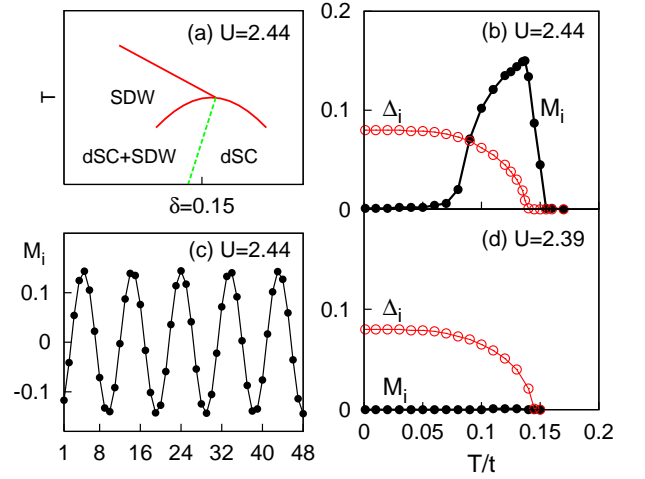}} \caption[*] {(a) The phase
diagram for $U=2.44$. (c) The projection of the staggered magnetization
$M_{\bf i}$ at $T=0.84T_N$, where $M_{\bf i}=(-1)^{\bf i}(n_{{\bf
i}\uparrow}-n_{{\bf i}\downarrow})$. The size of the unit cell is $N_x\times
N_y = 48 \times 24$. (b) and (d) show the temperature dependence of the $d$SC
(open circle) and the maximum value of $M_{\bf i}$ (solid circle) for
$U=2.44$ and $U=2.39$, respectively. The value of the $d$SC order parameter
$\Delta_{\bf i}=\frac{1}{4}(\Delta_{{\bf i}+\hat{x}}+\Delta_{{\bf
i}-\hat{x}}-\Delta_{{\bf i}+\hat{y}}-\Delta_{{\bf i}-\hat{y}})$ is measured
in the unit of $t$.}
\end{figure}

In Fig. 1(a), we use $ U= 2.44$ and $V=1.0$ to approximately reproduce the
phase diagram of Inaba $et\; al.$ \cite{inaba257} based on a mean field
approach of the $t-J$ model, there the contribution from the AF order was
explicitly considered. We shall apply this phase diagram to understand the
experimental observations in Ref. \cite{vershinin303}. It is important to
notice that the curve (dashed line) separating the region of the $d$SC and
the coexistence region of $d$SC plus SDW in the phase diagram has a positive
slop as the doping ( $\delta$) increases. The detailed structure of the SDW
in the underdoped region is not the main focus of the present paper and will
not be presented here. From the phase diagram, if $\delta=0.15$ is chosen
near the optimal doping or in the slightly underdoped region, one can easily
see that at low temperature $T$, our system is in the phase of $d$SC. When
$T$ increases, the SDW order emerges and coexists with the $d$SC. At very
higher $T$, the $d$SC disappears and only SDW order survives. These
temperature dependencies can be understood from Fig. 1(b) where the $d$SC
order parameter at site ${\bf i}$ and the maximum value of the staggered
magnetization $M_{\bf i}$ are plotted as functions of temperature $T/t$. From
Fig. 1(b), it is straightforward to show that only $d$SC exists and the AF
order is completely suppressed at low $T$. As $T>0.06t$, the stripe-modulated
AF or the SDW order incipit, and both of them could persist above the BCS
transition temperature $T_c^{BCS}=0.14t$. At $T>T_c^{BCS}$, the staggered
magnetization decreases rapidly to zero at the Neel's temperature
$T_N=0.155t$. In Fig. 1(c) we show that the projection of the $y$-oriented
stripe modulation in the AF order (or SDW) at $T=0.84 T_N < T_c^{BCS}$ along
$x$-axis, where the SDW has mixed periodicity $10a-9a-10a-9a-10a$ over a $48a
\times 24a$ lattice. The periodicity of the SDW seems not sensitive to $T$ as
long as the SDW order is in presence. In Fig. 1(d), we make similar plots as
those in Fig. 1(b) but with a smaller $U=2.39$. In this case the AF order is
completely suppressed and our system is in the state of pure $d$SC at all
temperatures. This implies that the phase diagram for $U=2.39$ should be
similar to that in Fig. 1(a) except the phase boundary (dashed line) is
pushed toward the lower doped region. In both figures 1(b) and 1(d), the
$d$SC order parameter as a function of $T$ appears to have the BCS-like
behavior. Besides, the $T_c^{BCS}$ in the case for $U=2.39$ [Fig. 1(d)] is
slightly larger than the one for $U=2.44$ [Fig. 1(b)]. This is because the
appearance of SDW in Fig. 1(b) at higher $T$ also suppresses the $d$SC. It
needs to point out here that a phase diagram of the same model was previously
studied by Martin $et\; al.$ \cite{martin56} using a very larger $U$ on a
much smaller lattice ($17a \times 10a$). The curve separating the region of
$d$SC and that of $d$SC plus SDW in their phase diagram has a negative slop.
And that would yield the conclusion that if the system is in pure $d$SC state
at low temperature, then it is always in $d$SC state at a temperature up to
$T <T_c^{BCS}$ and the SDW order never shows up. This is very different from
the present situation. In the following we show that the results in Figs.
1(b) and 1(c) can be applied to understand various features observed in the
STM experiment \cite{vershinin303}.

At finite temperature, what the STM measures is the LDTC which has the
following definition
\begin{eqnarray}
G_{\bf i}(E)_T \equiv  \left. \frac{dI_{\bf i}}{dE}\right|_T = A\int
\rho_{\bf i}(E')_T \biggl[-\frac{d}{dE}f(E'-E)\biggr] dE'
\end{eqnarray}
, where
\begin{eqnarray}
\rho_{\bf i}(E)_T &=&-\frac{1}{M_x M_y} \sum_{n,{\bf k}}^{2N} \biggl[  
\left| {\bf u}_{\bf i}^{n,{\bf k}} \right|^2 f'(E_{n,{\bf k}}-E) \nonumber
\\  &&+\left| {\bf v}_{\bf i}^{n,{\bf k}} \right|^2 f'(E_{n,{\bf k}} + E)
\biggr]
\end{eqnarray}
is the LDOS, and $A$ is proportional to the square of the tunnelling matrix
element. The summation in $\rho_{\bf i}(E)_T$ is averaged over $M_x \times
M_y$ wavevectors in first Brillouin Zone.

In Fig. 2, the normalized LDOS (a) and LDTC (b) at the sites with the
maximum staggered magnetization ($M_{\bf i}$) as functions of the bias
energy $E$ are presented from low to high temperatures. It is clear
from Eq. (4) that the LDTC reduces to LDOS at low $T$. But at higher $T$,
the behavior of the LDTC is rather different from that of the LDOS. For
example, the characteristic temperature $T_c^{bulk}$, where the coherent
peaks of the superconductivity begin to flatten out, in the LDTC is
considerably lower than that in the LDOS. One usually define the gap of the
quasiparticles as the separation between the coherent peaks or the width of
the dip or depression in the LDOS. While the gap in the LDOS [Fig. 2(a)]
appears roughly to be a constant from low $T$ up to $T_N$, the "gap" in
the LDTC
[Fig. 2(b)] increases progressively from low $T$ to high $T$. The increment
in the magnitude of "gap" as $T$ raises has been indeed observed by
experimental measurements \cite{timusk62, xuan0107, pan}. It should be
emphasized here that the "gap" in the LDTC or observed directly from the
STM experiments is the result of the convolution in Eq. (4) and may not be
the real gap of the system. In the temperature range of
$T_c^{bulk}<T<T_N$ as shown in Fig. 2, the system appears to be in the
"pseudogap" region according to the characterization from STM experiments
\cite{renner80, kugler86}. Even though the effect due to the phase
fluctuations on the $d$SC order parameter \cite{emery374, franz58} has not
been taken into account. In this region, the Fermi surface in our theory is
everywhere gaped while the ARPES experiment \cite{norman392} indicates that
the gaps occur only near ($\pm\pi, 0$) and ($0, \pm\pi$). This difficulty
so far has not been understood.

\begin{figure}[t]
\centerline{\epsfxsize=8.0cm\epsfbox{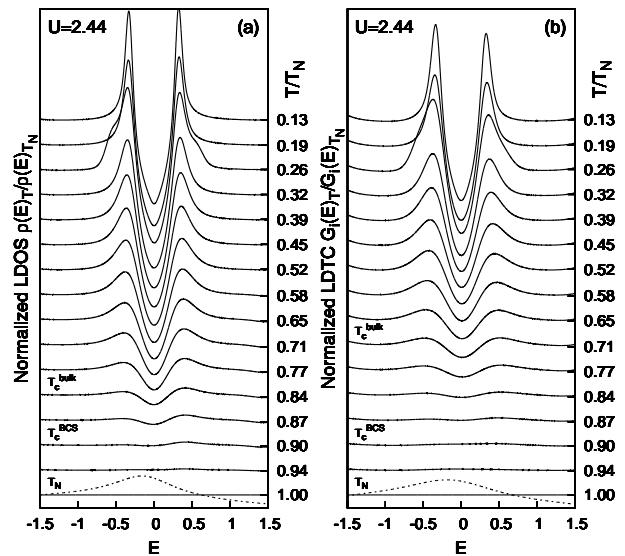}} \caption[*]{Temperature
dependence of the normalized LDOS ($\rho_{\bf i}(E)_T/\rho_{\bf i}(E)_{T_N}$)
and the normalized LDTC ($G_{\bf i}(E)_T/G_{\bf i}(E)_{T_N}$), here
$T_N=0.155t$. The LDOS and LDTC at $T_N$ as a function of E is represented by
the dashed line. This representative set of spectra was shifted vertically
for clarity. The wavevectors in first Brillouin Zone are $M_x\times
M_y=24\times 12$.}
\end{figure}

It is well established that associated with the SDW shown in Fig. 1 (c),
there are stripe-modulations in the $d$SC order and the charge density. The
stripe modulated charge order can also be refereed to the charge density
wave (CDW). In the top two graphs of the left panel in Fig. 3, the
projections of the $y$-oriented stripe modulations in the $d$SC order and
the charge density are plotted along the $x$-axis at $T=0.84 T_N (<
T_c^{BCS})$. Here the spatial distributions of the $d$SC and the charge
order are only slightly modulated by the stripe structure, and the $d$SC
and the SDW orders coexist in real space. This feature is very different
from the case in Ref. \cite{martin56}, where the $d$SC order is practically
suppressed to zero in the spatial regions where the AF order is in
presence. The stripe modulation also appears in the LDTC (or LDOS) as shown
in the bottom graph of the left panel. The stripe modulations displaced in
the left panel have the same mixed periodicity $5a-5a-4a-5a-5a$ over a
$24a$ lattice along $x$-axis. This indicates that the system is trying to
establish a periodicity incommensurate with the underline lattice, but
fails to do so because the calculation is performed in a discrete and
finite lattice, not in a continuum. It is also straightforward to show that
the same mixed periodicity still remains in the CDW and LDTC (or LDOS) even
$T$ is in the region of $T_c^{BCS} < T < T_N$, where the superconducting
order parameter vanishes and the SDW order is still in presence. As it will
be shown below that an average periodicity $4.8a$ for the stripe modulation
could be assigned in the present case. The $x$- and $y$- oriented stripe
modulations are degenerate in energy, and it is thus possible for the both
$x$- and $y$- oriented stripe-modulations to show up either in the time
duration when the experiment is performed or due to the proximity effect
\cite{howald67} between neighboring domains with stripes of different
orientations. As a result, checkerboard pattern could be observed.

\begin{figure}[t]
\centerline{\epsfxsize=8.0cm\epsfbox{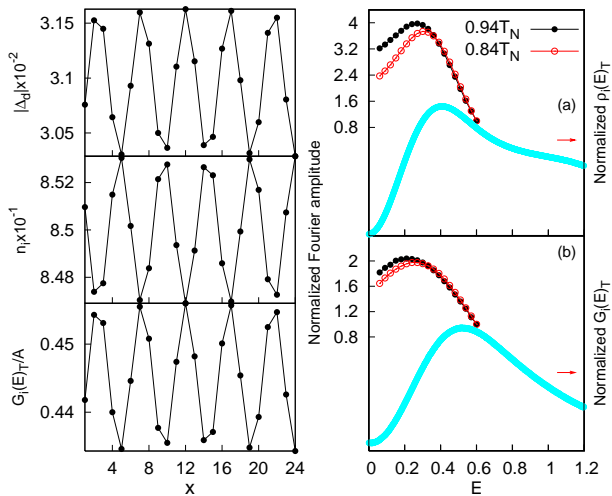}} \caption[*]{Left panel: From
top to bottom, the projection of the $d$SC order, charge density, and LDTC at
$T=0.84T_N$. Right panel: The energy evolution of the normalized Fourier
amplitude of the LDOS (a) and LDTC (b) images for $T=0.94T_N$ (solid circle)
and $T=0.86T_N$ (open circle) at $q_x=0.417(\pi/a)$. The solid curve
represents the normalized LDOS (a) and LDTC (b) as a function of $E$ at
$T=0.84T_N$. The wavevectors which have been used to calculate the LDOS and
LDTC images are $M_x\times M_y=6\times 6$ in first Brillouin Zone.}
\end{figure}

Furthermore, in order to get a detailed understanding of the periodicity of
the modulation in the LDOS (or LDTC) image, we perform the Fourier
transform of the LDOS (or LDTC) image in the temperature range where the
SDW order is in existence.
\begin{eqnarray}
\rho_{\bf q}(E)_T &=& \frac{1}{\sqrt{N_x N_y}} \sum_{\bf i} exp(i\;{\bf
q}\cdot {\bf r_i})\cdot \rho_{\bf i}(E)_T\;.
\end{eqnarray}
When the above quantity is plotted against $q_x$, a sharp peak occurs around
$q_x=0.417(\pi/a)=2\pi/4.8a$, indicating an average "periodicity" $4.8a$ in
real space. In Fig. 3(a) and 3(b), the solid curves at the bottom represent,
respectively, the normalized LDOS and the normalized LDTC as functions of the
bias energy at $T=0.84T_N$. The curves at the top made of solid/open circles
show the bias energy dependencies of the Fourier amplitudes of the LDOS and
LDTC images at $q_x=2\pi/4.8a$ for two different temperatures normalized by
their values at $E=0.6t$. The Fourier amplitude of the LDOS first reaches a
peak and then drops somewhat rapidly as the bias decreases to zero. Near the
zero bias, the result at $T=0.84T_N (<T_c^{BCS})$ dips more than the result
at $T=0.94T_N (>T_c^{BCS})$ in the pseudogap region. On the other hand, the
Fourier amplitude of the LDTC as functions of the bias energy at these two
temperatures differ very little, and they drops only slightly after reaching
a broader maximum from higher bias. This feature in the LDTC seems to be in
better agreement with the STM measurements \cite{vershinin303} as compared
with that of the LDOS.

In conclusion, the temperature dependencies of the LDTC and LDOS in a cuprate
superconductor with the competing AF order have been investigated in the
present paper. According to our calculations based on the phase diagram in
Fig. 1(a) near the optimal doping, there is no signature of the charge order
in the LDOS and the LDTC at low $T$. When the SDW order appears, both of the
LDOS and the LDTC exhibit the same CDW like modulation with an average
periodicity $4.8a$ at $T$ below and above $T_c^{BCS}$. We also calculate the
Fourier amplitude of LDOS and LDTC at ${\bf q}=(2\pi/4.8a,0)$ and their bias
energy dependencies in the "pseudogap" region as shown in Figs. 3(a) and
3(b), respectively. All the features in the LDTC are in good agreement with
those observed by the STM experiments \cite{vershinin303}. Finally we point
out that the "gap" of the quasiparticles at higher $T$ obtained directly from
the LDTC or the STM experiments does not correspond to the real gap of the
system in cuprate superconductors.

${\bf Acknowledgements}$: We thank S.H. Pan, J.X. Zhu and Q.
Yuan for useful comments and suggestions. This work is supported
by the Texas Center for Superconductivity and Advanced Material
at the University of Houston, and by a grant from the Robert A. Welch
Foundation.

\end{document}